\documentclass[11pt,preprint2]{aastex}

\usepackage{color}

\slugcomment{}

\shorttitle{Chromospheric and Coronal Observations of Flares with HMI}
\shortauthors{Mart\'inez Oliveros et al.}

\begin{document}


\title{\MakeUppercase{Chromospheric and Coronal Observations of Solar Flares with the Helioseismic and Magnetic Imager}}

\author{\textsc {Juan-Carlos Mart{\' i}nez Oliveros$^{1}$, S{\" a}m Krucker$^{1,2}$, Hugh~S. Hudson$^{1,3}$, \\ Pascal Saint-Hilaire$^{1}$, 
Hazel Bain$^{1}$, Charles Lindsey$^{4}$, Rick Bogart$^5$, Sebastien Couvidat$^5$, Phil Scherrer$^5$ and Jesper Schou$^6$}} 
\affil{$^1$Space Sciences Laboratory, UC Berkeley, Berkeley, CA 94720, USA\\
$^2$Institute of 4D Technologies, School of Engineering, University of Applied Sciences\\ North Western Switzerland, CH-5210 Windisch, Switzerland\\
$^3$School of Physics and Astronomy, University of Glasgow, Glasgow G12 8QQ, UK \\
$^4$North West Research Associates, CORA Division, Boulder, CO 80301, USA\\
$^5$W.~W. Hansen Experimental Physics Laboratory, Stanford University, Stanford, CA 94305, USA\\
$^6$Max-Planck-Institut f{\"u}r Sonnensystemforschung, Justus-von-Liebig-Weg 3, D-37077 G{\"o}ttingen, Germany}



\begin{abstract}
We report observations of white-light ejecta in the low corona, for two X-class flares on the 2013~May~13, using
data from the Helioseismic and Magnetic Imager (HMI) of the \textit{Solar Dynamics Observatory}.
At least two distinct kinds of sources appeared (chromospheric and coronal), in the early and later phases of flare development, in addition to the white-light footpoint sources commonly observed in the lower atmosphere.
The gradual emissions have a clear identification with the classical loop-prominence system, but are brighter than expected and possibly seen here in the continuum rather than line emission.
We find the HMI flux exceeds the radio/X-ray interpolation of the bremsstrahlung produced in the flare soft X-ray sources by at least one order of magnitude.
This implies the participation of cooler sources that can produce free-bound continua and possibly line emission detectable by HMI.
One of the early sources dynamically resembles ``coronal rain'', appearing at a maximum apparent height and moving toward the photosphere at an apparent constant projected speed of 134~$\pm$~8~$\mathrm{km~s^{-1}}$.
Not much literature exists on the detection of optical continuum sources above the limb of the Sun by non-coronagraphic instruments, and these observations have potential implications for our basic understanding of flare development, since visible observations can in principle provide high spatial and temporal resolution.
\end{abstract}



\section{Introduction}

Only extremely rare solar flares have produced white-light ejecta detectable above the limb of the Sun and imaged without the aid of a coronagraph.
To our knowledge, the modern literature contains only three such events: SOL1980-06-21 (X2.6), an event visually observed by J. Harvey and T. Duvall \citep{1983SoPh...86..123H,1990ApJS...73..213C}, SOL1989-08-14 \citep[$\sim$X20; ][]{1992PASJ...44...55H} and SOL2003-11-04  \citep[$>$X17;][]{2004AAS...204.0213L}.
In all cases the flares occurred at heliographic positions near the limb, and the emissions were observed in the lower atmosphere (below 1.03~$R_\odot$ elongation).
In many more cases white-light ejecta continue away from the Sun in the form of coronal mass ejections (CMEs), observed at substantial elongations and with specialized telescopes (e.g. coronagraphs), but these sources clearly have different morphologies from the sources mentioned above.
In this Letter we make an initial report of chromospheric and coronal emission sources observed by the \textit{Solar Dynamics Observatory}/Helioseismic and Magnetic Imager (\textit{SDO}/HMI) for two flares in 2013 May: SOL2013-05-13T02:17 (X1.7) and SOL2013-05-13T16:01 (X2.8), each associated with a CME and a meter-wavelength type II radio burst.
The new HMI  sources appear in an image annulus extending about 25$''$ above the limb, and this report provides a first scientific use of such data.
We compare the detailed observations of these sources with X-ray observations from the \textit{Reuven Ramaty High-Energy Solar Spectroscopic Imager} \citep[\textit{RHESSI};][]{2002SoPh..210....3L}, and with other \textit{SDO} observations.

The HMI instrument contains two cameras (front and side) designed for on-disk observations of the whole Sun, primarily for helioseismology and the characterization of the photospheric magnetic field, and for these purposes it makes high-resolution filtergrams of the photospheric Fe {\sc i} absorption line at 6173.3~\AA. 
HMI observes the line via  six pass bands spanning a range of about 345~m\AA~around the target line and it also obtains full Stokes profiles \citep{2012SoPh..275..229S}. The ``front camera" standard data are created from a combination of sequences of 12 distinct images (sex filters and two polarizations each called a filtergram), while the ``side camera" data are  reconstructed from observations with the same filters and the following polarizations: I+/--U, I+/--Q, and I+/--V. The individual wavelengths (and polarization settings) for these line profiles are not observed simultaneously, but in a programmed sequence extending over 45~s and 135~s frames for the HMI front and side cameras, respectively,  and therefore have limitations when rapid transients occur \citep[e.g.,][]{2013SoPh..tmp..209M}. The resulting images are cropped and later during the process, truncated at about 25$''$ above the limb to generate the standard HMI observables (intensity, velocity and magnetic field). The observations we report here, though unambiguous photometrically, thus occur in a parameter space not optimized by the design of the telescope.
To help understand the signatures, we also study the individual filtergrams, which cycle through sets of polarizers and spectral filters. In this initial Letter we emphasize the source morphology and do not attempt serious quantitative analysis of, for example, the polarization signatures which will be presented in a subsequent Letter.

\section{The HMI 6173.3A narrow band phenomena}\label{sec:hmi}

\begin{figure*}[htbp]
\centering
   \includegraphics[width=0.95\textwidth]{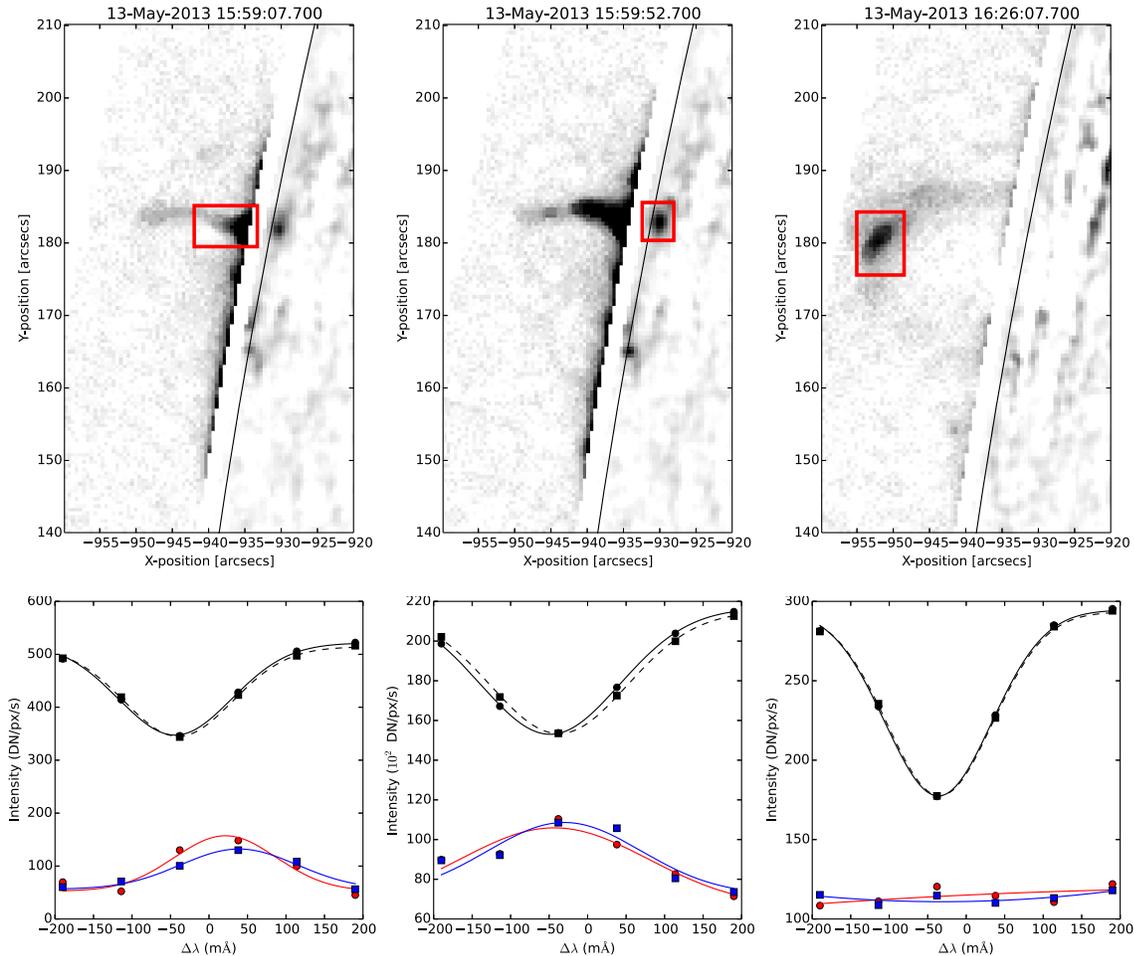}
    \caption{HMI intensity images at three times during SOL2013-05-13T16:01, and six-point spectra averaged over the boxed regions. Left, an early low-altitude jet-like source; middle, a northern footpoint (here with a different scaling); right, the late-phase loop-like source. The image panels have a simple contrast enhancement (factor of 80) in the corona, beginning a few pixels above the limb itself (black line). The mean spectra refer to selected pixel groupings and have the simultaneously measured background levels subtracted.
     The averaged spectra derived from the left circular polarization  (LCP) data is shown by the red circles and the red line shows the best Gaussian fit to the data; similarly blue shows the right circular polarization (RCP). 
  In each spectrum the black solid and dashed curves shows the reference spectrum from its background region for each polarization, consisting of the non-flaring part of the annulus that encompasses the source pixels.  
     }  
\label{fig:image_jc}
\end{figure*}

During the SOL2013-05-13T02:17 and SOL2013-05-13T16:00 events, the HMI front camera (the standard HMI intensity) shows several features above the limb: short-lived jet-like early sources associated with the impulsive phase of the flare (Figure~\ref{fig:image_jc}: red box in left frame),  long-lived loop-like later sources (Figure~\ref{fig:image_jc}: red box in right frame),  and an enhanced glow along the limb from the chromosphere above the flaring active region. Although both events show these distinct features we will focus on the SOL2013-05-13T16:00 event as it is the best observational case.

\begin{figure*}[htb]
\centering
   \includegraphics[width=\textwidth]{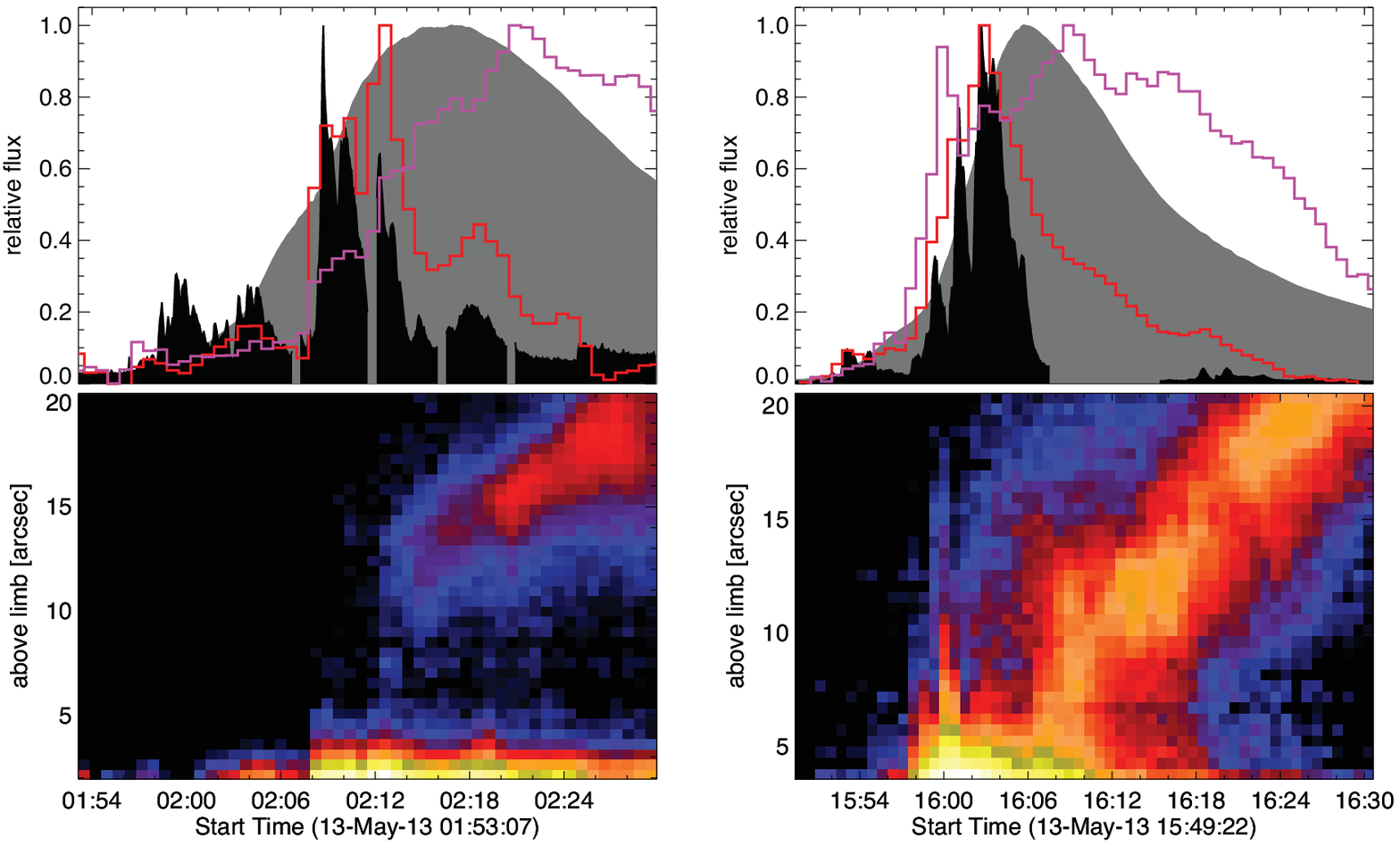}
    \caption{Time series plots for the two flares studied here.
    The upper panels show the \textit{GOES} long-wavelength channel (gray fill), the \textit{RHESSI} 30\,--\,100~keV flux (black fill, with some gaps), and two histograms with HMI summed intensities  in the early source region (red) and in a field of view containing all of the observed phenomena (purple), both early and gradual.
    The light curves show excess fluxes, relative to the start time, normalized to their maxima.
    The lower panels show height vs. time information for HMI; note the spike at about 16:00~UT in the lower-right panel, which appears in only a few 45~s data frames.
    The limb reference position is at elongation 949$''$.06.
   }  
\label{fig:events}
\end{figure*}

Figure~\ref{fig:events} shows the coronal sources as height versus time plots and time series, along with \textit{GOES} and \textit{RHESSI} X-ray data.
The late sources move radially outward for some tens of minutes and eventually vanish outside the truncated radius of the HMI  images; the short-lived brightenings during the impulsive phase remain at lower altitudes.
The sources are generally fainter than the background brightness due to stray light from the solar disk.
In this work we reduce the background by subtracting a preflare image,\footnote{In principle we could do this via the point-spread function of HMI, but for this analysis we have chosen to do it via simple image subtraction.} which minimizes the effects of scattered light from the solar atmosphere and from within the instrument itself.
We find the background light level to be quite stable and to have a lower time series fluctuation level than the on-disk images.
The image brightness in the annulus is about 0.5\%--2\% of the disk-center image brightness, but the rms fluctuation in a difference
image, which limits the sensitivity, is typically only about 0.02\%.

The significance of the HMI detection of these sources depends on identifying the emission mechanisms and in understanding the associations of the HMI sources  with phenomena already identified at other wavelengths.
The sources we see are relatively bright with a similar evolution and morphology as the white-light flare observed by \citet{1992PASJ...44...55H}, and may thus entail substantial mass, energy, or momentum.
In principle the HMI pass band could have a strong contribution from emission in the Fe~{\sc i}~6173.3\AA~line for which it was designed, but in these rarely observed sources other mechanisms may dominate because of the unusual physical conditions.
We suggest two other obvious candidate emission mechanisms: Thomson scattering from free electrons in the material ejected from the lower atmosphere, and a combination of free\,--\,free, free\,--\,bound and bound\,--\,bound emissions in the visible range as seen in loop-prominence systems \citep[e.g.,][]{1960BAICz..11..165K,1965ApJ...141..519J}.
The latter mainly consists of free\,--\,free continuum for high temperatures, but  includes bright contributions from free\,--\,bound and bound\,--\,bound transitions at lower temperatures.
If these sources have high densities, Thomson scattering becomes less likely because of its linear dependence on $n_e$, while the different components of the recombination radiation are proportional to the product $n_e n_i \sim n_e^2$.
The morphology of the gradual-phase sources strongly suggests the  coronal mass cycle of a solar flare\footnote{The coronal mass cycle of a solar flare refers to the well-known injection/evaporation/cooling cycle observed during flares.}; this results in the formation of a loop-prominence system, known from H$\alpha$ observations to require high densities in some major flares \citep[e.g.,][]{1972ARA&A..10....1S}.

\begin{figure*}[htbp]
\centering
   \includegraphics[width=0.49\textwidth]{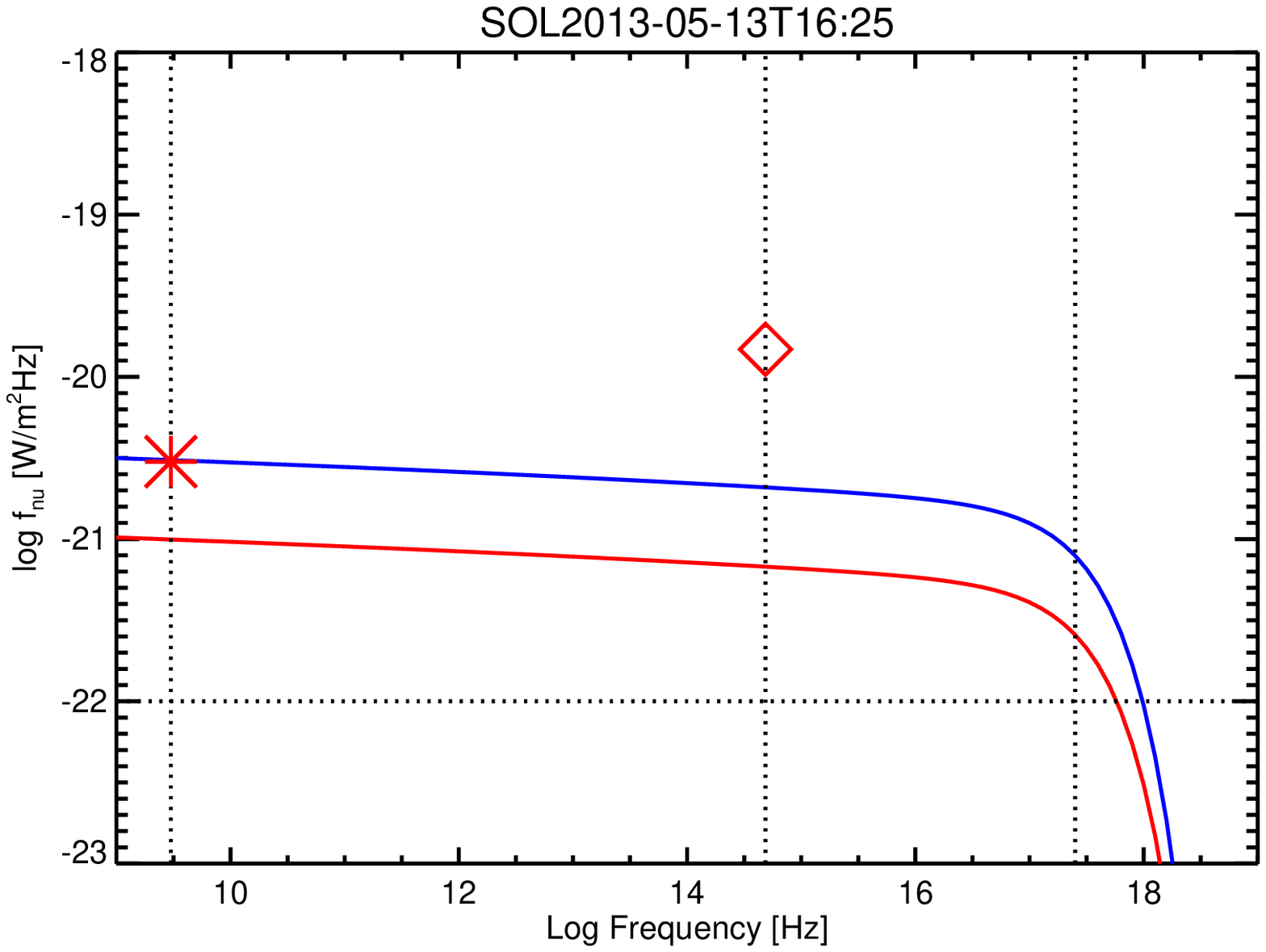}
   \includegraphics[width=0.49\textwidth]{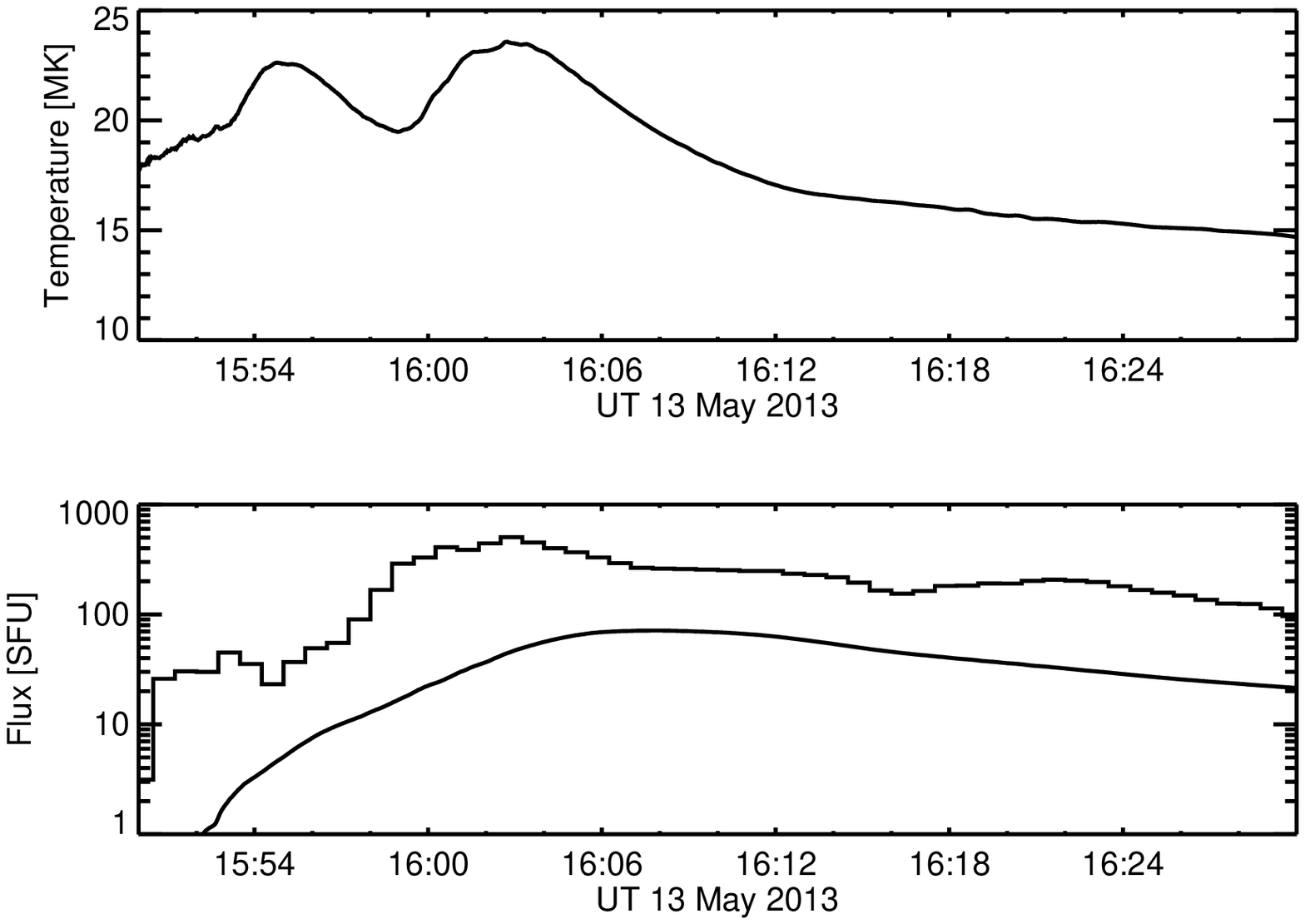}
    \caption{Left, spectral comparison of free\,--\,free emission levels at microwave and HMI wavelengths, as predicted by the \textit{RHESSI} thermal parameters (red line) and those obtained from \textit{GOES} (blue line), in a snapshot at 16:25~UT (\textit{GOES}: emission measure $4.1 \times 10^{49}$~cm$^{-3}$, temperature 15.1~MK; \textit{RHESSI}: emission measure $1.4 \times 10^{49}$~cm$^{-3}$, temperature 17.5~MK).
    At this temperature free\,--\,bound emission plays no role at long wavelengths.
    The vertical lines show the frequencies of 1~cm, 6173~\AA, and 1~keV.
     The total HMI flux (red diamond) and RSTN 15.4~GHz (red star) flux densities are from 16:25~UT.
    Right, the time variations of the predicted free\,--\,free flux (smooth line), and observed HMI fluxes (histogram).
   }  
\label{fig:freefree}
\end{figure*} 

The free\,--\,free continuum underpins the soft X-ray flare emission and the ``post-burst increase'' microwave emission \citep[][]{1972SoPh...23..155H,1972SoPh...23..169S,1996A&A...314..947P}.
This component must be present at visible wavelengths from the same coronal sources that produce the soft X-ray and microwave fluxes, since the corona is optically thin at the HMI observing wavelengths. As shown in Figure~\ref{fig:freefree}, the HMI flux in the gradual-phase source significantly exceeds that expected from free\,--\,free emission
with the emission measure and temperature of the soft X-ray sources. The free\,--\,free contribution was calculated using the formula \citep[e.g.,][]{1980afcp.book.....L} $$f_{\mathrm{HMI}} = 3.5 \times 10^{-68} \bar{g} \mathrm{EM}/\sqrt{T}\ \ \mathrm{W/m^2Hz\ ;}$$ here $\bar{g}$ is the mean Gaunt factor for free\,--\,free emission, $\mathrm{EM}$ the emission measure ($n_e n_i V$) in cm$^{-3}$, and $T$ the temperature in K.
This could imply an additional emission measure from free\,--\,free emission at temperatures not detectable in the X-ray band \citep{1995SoPh..160..181C}, a contribution from Thomson scattering, or the change of the ionization state of the observed material as it cools catastrophically \citep{1965ApJ...142..531F, 1971SoPh...19...86G}. In principle, using the spectropolarimetric capabilities of HMI we can directly determine the presence of a Thomson-scattered component via its linear polarization \citep{1930ZA......1..209M}. In this Letter we do not attempt a detailed interpretation of the spectral or polarization data of the HMI coronal sources, but we do refer to them for qualitative information.

Figure~\ref{fig:image_jc} shows spatially averaged spectra for three areas in SOL2013-05-13T16:01.
These spectra are differences relative to non-flare regions at the same radial coordinates; the background spectra show the absorption line clearly.
Instead of an absorption line, the footpoint and chromospheric sources reveal the line to be in emission, whereas the loop-like coronal source at a later time does not show either emission or absorption.
The HMI observations of these low-corona sources thus require either a source of continuum, which could include free\,--\,bound components of the recombination radiation, or a pseudo-continuum formed from highly Doppler-shifted line emissions.
The Doppler range of HMI only corresponds to $\pm$6.5~$\mathrm{km~s^{-1}}$, much less (for example) than the apparent transverse speed we measure with the HMI images in the early source of SOL2013-05-13R16:01; since we know little about the geometry of the source or its line-of-sight velocity distribution, we cannot rule out the possibility of a line contribution to the signal, and this could even include the Fe~{\sc i} line itself in emission, although this state should not be abundant in the corona.
In any case the free\,--\,free extrapolation fails by about an order of magnitude to explain the observed HMI flux density in the coronal emission region.

The early sources in SOL2013-05-15T16:01, in particular, appear to be somewhat different from the later loop-like sources.
A movie representation of the source at about 16:00~UT has the appearance of coronal rain. The emission appears first at its maximum height and then moves downward; we estimate the transverse velocity from the apparent motion to be 134~$\pm$~8~$\mathrm{km~s^{-1}}$.
We have characterized this motion by constructing a movie combining the individual filtergrams from both HMI cameras.
In this manner HMI crudely achieves a cadence of 1.875~s, since the two cameras have been set to record temporally interleaved images. 
At about the same time the Atmospheric Imaging Assembly (AIA) movies show outward motion in an adjacent region that reaches far beyond the outer edge of the HMI image limit, in what is no doubt a true jet.
The HMI emission appears near the middle of the annulus extending about 25$''$ above the limb, at about a 10~Mm altitude, and also appears in the AIA EUV bands (Figure~\ref{fig:impulsive}).
Its patchy structure as seen in all spectral ranges also suggests the thermal instability of the coronal rain phenomenon \citep[e.g.,][]{1972ARA&A..10....1S}.
The likely interpretation of the HMI source is then that it represents a short-lived compact loop system associated with jet activity in the impulsive phase and that HMI does indeed record coronal rain appearing in the final stages of its thermal collapse.
The rapid evolution to the cool phase makes sense if the loops are compact and dense to begin with; note that the \textit{GOES} temperature variation (right panel of Figure~\ref{fig:freefree}) shows an early flare episode not readily seen in the emission measure.
We are unaware of detailed modeling of these stages of the thermal collapse of cooling flare loops, but it seems reasonable that the Fe~{\sc i} line itself might go into emission during this time.

\section{\textit{RHESSI} X-ray observations}\label{sec:rhessi}

\begin{figure*}[htb]
\centering
   \includegraphics[width=\textwidth]{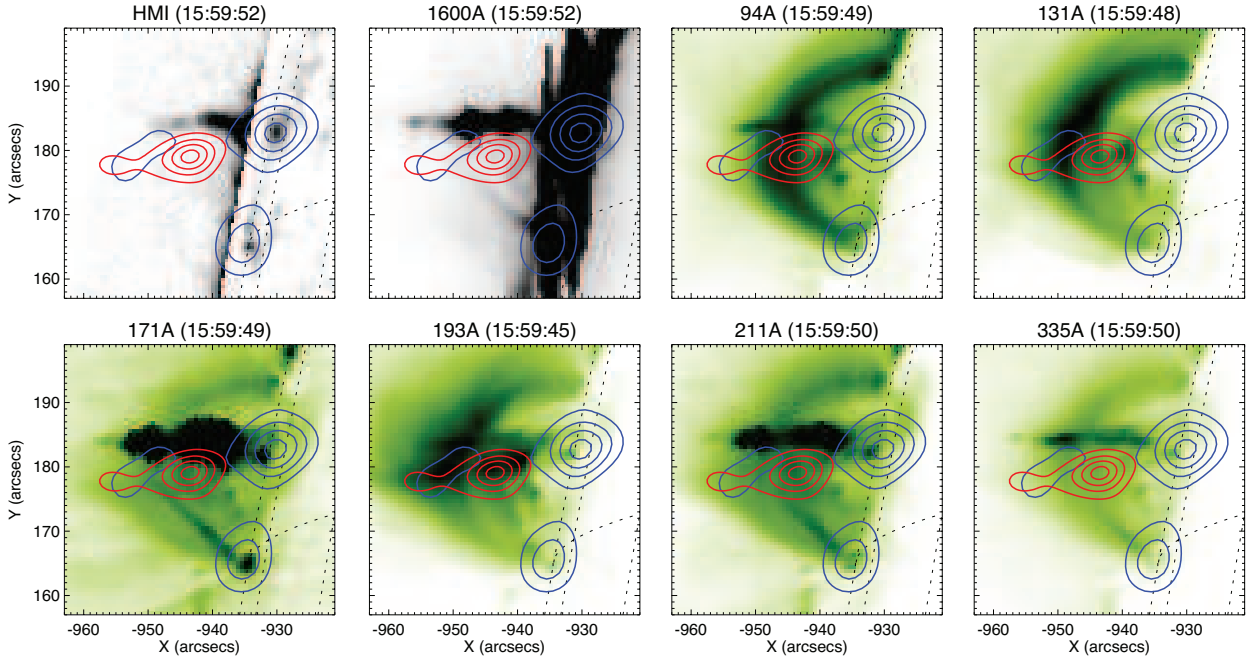}
    \caption{Impulsive-phase sources, as seen in gray-scale images from HMI (upper left) and seven AIA bands.
    The contours show \textit{RHESSI} sources: red, 12\,--\,25~keV; blue, 30\,--\,100~keV.
    The hard X-ray and white-light footpoint sources align as shown with no adjustment needed, even for the roll coordinate.
    Note that the prominent loop apparently connecting to one of the flare footpoints (e.g., in the 94~\AA~band) was present before the start of the main HXR bursts (15:59UT) and thus not directly related to it.
    The \textit{RHESSI} soft X-ray sources lie to one side of the jet-like feature, but note that the 193~\AA~pass band shows a coincident component that would be consistent with the well-known Fe~{\sc xxiv} response in this filter.
   }  
\label{fig:impulsive}
\end{figure*}

The X-ray observations provide a key link between known flare phenomenology and the phenomena described here.
Fortunately \textit{RHESSI} had excellent observations of these two events (Figure~\ref{fig:events}), and we review briefly the hard and soft X-ray sources as they relate spatially and temporally with the impulsive and gradual sources observed by HMI. Figure~\ref{fig:impulsive} shows the impulsive-phase development of SOL2013-05-13T16:01 as seen by the AIA in the EUV,  and by \textit{RHESSI} in soft and hard X-rays.
The footpoint hard X-ray sources are well-matched with the white-light emission patches (Figure~\ref{fig:impulsive} upper left frame).
The AIA EUV images (six panels as labeled) show foreground/background sources that appear to be loops in the active region resulting from early stages of the flare development.
The jet-like source originates near the northern footpoint, but has a problematic interpretation which we will discuss below.

The \textit{RHESSI} soft X-ray images (red contours in Figure~\ref{fig:impulsive}) reveal a source at a comparable height, but displaced laterally from the jet-like HMI source.
The AIA 193~\AA~pass band also shows a feature at this location, confirming that it has a hot component via the Fe~{\sc xxiv} contribution to this pass band. This hot source is not observed by HMI or the more sensitive AIA at the 1600~\AA~pass band. This situation is made more interesting by the fact that the HMI jet-like source does not have a hot counterpart observed by \textit{RHESSI} or AIA hot channels (\textit{RHESSI} in general cannot detect sources with temperatures below 8~MK). This suggests that the HMI emission is not the bremsstrahlung tail  from a hot source, but  free\,--\,bound emission from plasma at temperatures too low to emit X-ray radiation detectable by \textit{RHESSI}.
This possibility of low-temperature ejecta into closed-field structures in the impulsive phase of a flare does not fit naturally into our conventional understanding and hence may represent something new about flare development.
The \textit{RHESSI} hard X-ray contours, on the other hand, reflect the well-known footpoint sources and hint at a loop-top source as well.

\begin{figure*}[htbp]
\centering
   \includegraphics[width=\textwidth]{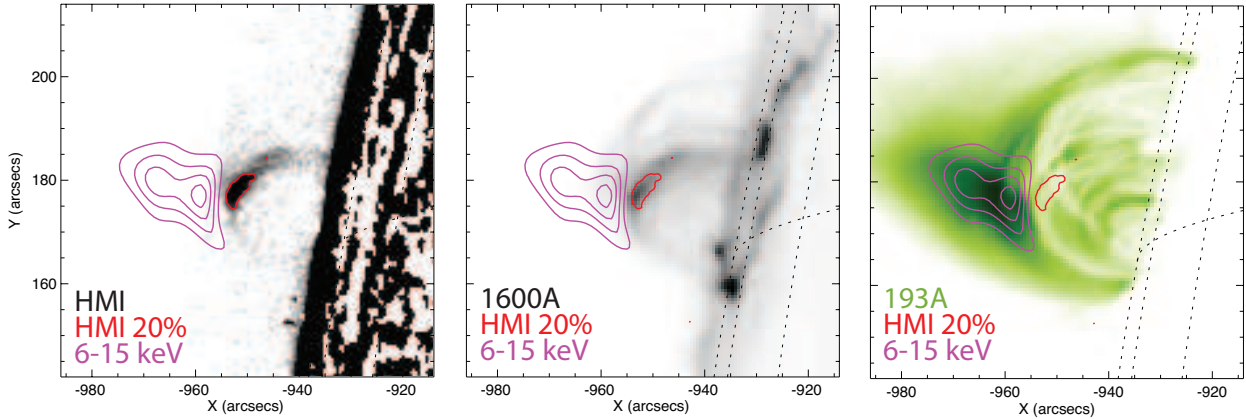}
    \caption{Gradual-phase sources, with image times WL: 16:25:22.7~(WL), 16:25:28.1 (1600~\AA), and 
16:25:21.5 (193~\AA). The HMI 193~\AA\, image shows an absorption feature just at the lower edge of the soft X-ray source, which itself must be optically thin.
   }  
\label{fig:gradual}
\end{figure*}

The HMI loop-like sources, observed during the gradual phase of the events, appear to be easier to interpret.
We see a close relationship with the loops seen in the AIA 1600~\AA~bands, but not directly in the EUV bands, where
most of the emission is from larger-scale structures (Figure~\ref{fig:gradual}).
This component and the coronal soft X-ray sources seen by \textit{RHESSI} are concentrated along the apex of an arcade of loops, whose axis roughly lies along the line of sight but with the northern footpoints noticeably displaced to the west, as judged from the HMI images.
They also suggest a cusp structure that reflects the open field lines created by the associated CME \citep[e.g.,][]{1993GeoRL..20.2785H}.

\section{Discussion} 

The discovery of these HMI features was made in the course of searches for white-light flare emission. The HMI observations of chromospheric and coronal emission sources surprised us, because they must be very bright in order to be visible against the normal instrumentally scattered background outside the white-light limb.
In SOL2013-05-13T16:00 the white-light footpoints are much brighter (1--2 decades) than the coronal sources, but the HMI data have an excellent dynamic range and image stability, so simple difference imaging immediately revealed these sources, and will probably reveal many other such flare features in other events.
We have found similar features in SOL2011-01-28T01:03 (M1.3) and SOL2012-01-27 (X1.3), for example.
These may be the first reports of such low-coronal features in white light from a general-purpose telescope in space, but they have been preceded by ground-based observations \citep{1992PASJ...44...55H,2004AAS...204.0213L}; accordingly we expect that many more flares will be detectable in this manner through the HMI data.
We have tentatively identified the HMI gradual-phase signatures with the well-understood phenomenon of the evaporation/loop-prominence/coronal rain ``coronal mass cycle'' of a flare \citep[e.g.,][]{1972ARA&A..10....1S}.
The HMI early sources in the second flare have some differences, but still fit  reasonably well into this standard scenario for the coronal mass cycle.
One outstanding question, though, is why HMI does not see the actual sources that \textit{RHESSI} sees;
we speculate that these sources are indeed there, but that the HMI source ``ignites'' and becomes detectably bright only when the thermal collapse happens at the lower altitude of the HMI source \citep{1971SoPh...19...86G}.
Note that we have not discussed the extended brightening along the limb at all in this Letter, nor have we attempted to decipher the
wavelength and polarization information from HMI, and so considerable important work remains to be done.

The steady apparent upward motions of the loop-like sources during the gradual phase of the events pose problems, since they do not seem to reflect a physical shrinkage of the magnetic structures during the dipolarization of the reconnected field volumes as suggested by the scenario of the coronal mass cycle \citep[e.g.][]{1987SoPh..108..237S,1996ApJ...459..330F}, but in this picture the density also decreases as the cooling proceeds \citep{1991A&A...241..197S}.
The cycle ends with the catastrophic cooling that terminates the shrinking motion, and so it is possible that this altitude systematically moves up with time; the implication of this development is that no appreciable amount of material remains in these terminally cooled loops.
The apparent upward motion has a projected speed of about 7~$\mathrm{km~s^{-1}}$, not inconsistent with typical speeds of arcade growth in loop prominence systems.

These observations have great diagnostic potential from the AIA imaging, which shows related chromospheric and coronal sources in all of its pass bands.
As Figure~\ref{fig:gradual} shows, these include loops seen in absorption, as recently discussed in detail by \cite{2013ApJ...772...71L} for absorption by prominences.
These absorbing loop features presumably contain material at chromospheric temperatures that is not a part of the hot condensation/coronal rain scenario, either unrelated foreground features or else flare loops in the final stages of cooling.
Such features absorb strongly  in the EUV, but should be transparent to both HMI continuum and \textit{GOES} soft X-radiation.
The absorbing features do not obviously affect the AIA 1600~\AA, as seen in the images at 16:25~UT in Figure~\ref{fig:gradual}.

We note that Nobeyama observations are available for SOL2013-05-13T02 and for the other two X-class flares in 2013 May; these data image the thermal ``post-burst increase'' sources in detail.
Future studies of these events can therefore compare an unprecedented collection of radio, visible, EUV, and X-ray data, which we feel will be especially interesting in understanding the end phase of the coronal mass cycle in major two-ribbon flares.

\section*{Acknowledgements}
This work was supported by NASA (Contract NAS5-98033 for \textit{RHESSI}) and by the Swiss National Science Foundation (200021-140308).
Some co-authors were supported by NASA Grant NAS5-02139 (HMI). The data used here are courtesy of NASA/\textit{SDO} and the HMI and AIA science teams. We thank the CHIANTI consortium \citep{1997A&AS..125..149D} for their excellent atomic-physics tools. This research made use of Sunpy, a community-developed Python package for Solar Physics \citep{2013SPD....44..136C} and SolarSoft, an analysis environment for solar physics \citep{1998SoPh..182..497F}.

\bibliographystyle{apj}

\end{document}